\documentclass{llncs}
\usepackage[utf8]{inputenc}
\usepackage[english]{babel}
\usepackage{amsmath}
\usepackage{amsfonts}
\usepackage{amssymb}
\usepackage{gauss}

\usepackage{blkarray}
\usepackage{tikz}
\usepackage{relsize}
\usepackage{numprint}
\usepackage[linesnumbered,noend,noline]{algorithm2e}

\def\clap#1{\hbox to0pt{\hss#1\hss}}

\begin{document}

\title{Good Pivots for Small Sparse Matrices}
\author{Manuel Kauers\thanks{M.K. was supported by the Austrian FWF grants F50-04, W1214-13, and P31571.}\orcidID{0000-0001-8641-6661} \and
  Jakob Moosbauer\thanks{J.M. was supported by the Land Oberösterreich through the LIT-AI Lab.}\orcidID{0000-0002-0634-4854}}
\institute{%
  Institute for Algebra, Johannes Kepler University, Linz, Austria\\\relax
  [manuel.kauers$|$jakob.moosbauer]@jku.at
}

\bibliographystyle{splncs04}

\maketitle

\begin{abstract}
  For sparse matrices up to size $8\times8$, we determine optimal choices for pivot
  selection in Gaussian elimination. It turns out that they are slightly better than
  the pivots chosen by a popular pivot selection strategy, so there is some room for
  improvement. We then create a pivot selection strategy using machine learning and
  find that it indeed leads to a small improvement compared to the classical strategy.
\end{abstract}

\section{Introduction}

It can be cumbersome to solve a sparse linear system with Gaussian elimination because
a poor choice of a pivot can have a dramatic effect on the sparsity. In the worst case,
we start with a fairly sparse matrix, and already after a small number of elimination
steps, we are faced with a dense matrix, for which continuing the elimination procedure
may be too costly. The principal goal of a linear system solver for sparse matrices is
therefore to maintain as much of the sparsity as possible, for as long as possible. To
achieve this goal, we can pay special attention to the pivot selection. A popular
pivot selection strategy which aims at maintaining the sparsity of a matrix is attributed
to Markowitz~\cite{Ma:TEFO,DE:DMFS}. It is based on the notion of ``fill-in'', which is defined as
the number of matrix entries that get affected when a particular element is chosen as
pivot. More precisely, for a matrix $A=((a_{i,j}))_{i,j=1}^{n,m}$, let $r_i$ be the
number of nonzero entries of the $i$th row ($i=1,\dots,n$) and $c_j$ be the number of
nonzero entries of the $j$th column ($j=1,\dots,m$). Then the fill-in associated to the
entry at position $(i,j)$ is defined as $(r_i-1)(c_j-1)$. Note that this is exactly the
number of cells into which something gets added when the entry at $(i,j)$ is chosen as
pivot: 
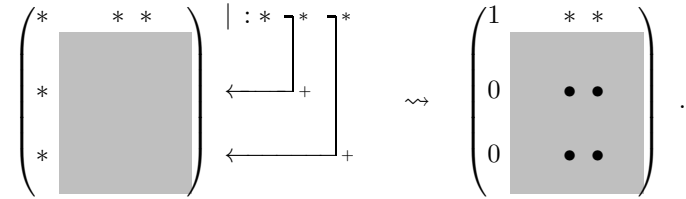
\[
 \begin{gmatrix}[p]
  \ast & \rule{1.8em}{0pt} & \ast & & \ast & \rule{1em}{0pt} \\
  \rule{0pt}{1.3em}        &      & &     &          \\
  \ast &                   &      & &     &          \\
  \rule{0pt}{0.9em}        &      & &     &          \\
  \ast &                   &      & &     &          \\
       & \smash{\rlap{\color{lightgray}\kern-.8em\rule{5em}{6.1em}}}
       &      & &     &         
 \rowops
  \mult0{:\ast}
  \add[\ast]02
  \add[\ast]04
 \end{gmatrix}
 \quad\leadsto\quad
 \begin{gmatrix}[p]
  1 & \rule{1.8em}{0pt} & \ast & & \ast & \rule{1em}{0pt} \\
  \rule{0pt}{1.3em}        & \smash{\raisebox{-4.8em}{\rlap{\color{lightgray}\kern-.8em\rule{5em}{6.1em}}}}     & &     &          \\
   0   &                   & \bullet     & & \bullet    &          \\
  \rule{0pt}{0.9em}        &      & &     &          \\
   0   &                   & \bullet     & & \bullet    &          \\
       & 
       &      & &     &         
 \end{gmatrix}.
 \]
 The selection strategy of Markowitz is to choose among the eligible candidates a pivot for
 which the fill-in is minimized.
 The strategy thus neglects that touching a cell that already is nonzero does not decrease
 the sparsity (it may in fact increase if we are lucky enough).
 
 It is known that finding a pivot that minimizes the number of new entries introduced during 
 the whole elimination process has been shown to be NP-complete by Yannakakis \cite{Ya:CTMF}.
 But how much does this matter? In other words: how close does the Markowitz pivot selection strategy
 get to the theoretical optimum? This is the first question we address in this paper. By an
 exhaustive search through all square matrices up to size $8\times8$ in an idealized setting,
 we have determined the pivot choices that minimize the total number of operations. As expected,
 it turns out that with the optimal pivot choice, the number of operations is indeed smaller
 than with the Markowitz strategy, albeit just by a small amount. This confirms the common
 experience that the Markowitz strategy is a good approach, especially since it also has the
 feature that it can be easily implemented and does not cost much.
 Nevertheless, there is some room for improvement, and the second question we address in this
 paper is how this room for improvement could possibly be exploited.
 We tried to do so by training a pivot selection strategy using machine learning. It turns out
 that the resulting neural network performs indeed a bit better than the Markowitz strategy, at
 least in the setting under consideration.

 Our study is limited to small matrices because determining the optimal pivots by exhaustive search
 is prohibitively expensive for larger matrices. It is clear that sparsity optimization for matrices
 of this size does not have any practical relevance. In fact, it may be argued that there are
 no nonzero sparse matrices of size $8\times 8$ at all, because each such matrix has at least 12.5\%
 nonzero entries, which is far more than the sparsity of matrices arising in many numerical applications.
 However, our results indicate that the gap between Markowitz criterion and the optimal choice can
 possibly be narrowed by adequate use of machine learning, and it may have some relevance for sparse
 matrices with symbolic entries for which the cost of arithmetic is so high that spending some additional
 time on searching for a better pivot may be justified. In future work, we will investigate whether
 the machine learning approach can also be used to construct a selection strategy for symbolic matrices
 of more realistic sizes.

 \section{Algebraic Setting}
 We will distinguish two kinds of matrix entries: $0$ (zero) and $\ast$ (nonzero), and we ignore the
 possibility of accidental cancellations, so we adopt the simplifying assumption that the sum of two
 nonzero elements is always nonzero.
 As coefficient domain, we therefore take the set $S=\{0,\ast\}$ together with addition and multiplication
 defined as follows:
 \[
 \begin{array}{c|c c}
   + & 0 & \ast\\
   \hline
   0 & 0 & \ast\\
   \ast & \ast & \ast\\
 \end{array}
 \qquad			
 \begin{array}{c|c c}
   \cdot & 0 & \ast\\
   \hline
   0 & 0 & 0\\
   \ast & 0 & \ast\\
 \end{array}.
 \]
 The operations we count are $\ast+\ast$ and $\ast\cdot\ast$, i.e., operations not involving zero.
 As we have primarily applications with symbolic matrices in mind
 (originating, e.g., from applications in symbolic summation~\cite{koutschan13}, 
 Gr\"obner bases computation~\cite{faugere99}, 
 or experimental mathematics~\cite{BR:CMT,CT:TBEP}), we do not worry about stability
 issues. 

 The exact number of operations depends not only on the choice of the pivot but also on how the
 elimination is performed. We consider two variants. In the first variant, we add a suitable
 multiple of the pivot row to all rows which have a nonzero entry in the pivot column:
 \[
 \begin{gmatrix}[p]
   a & b & 0 \\
   0 & c & d \\
   e & f & g
   \rowops
   \add[-e/a]02
 \end{gmatrix}
 \leadsto
 \begin{gmatrix}[p]
   a & b & 0 \\
   0 & c & d \\
   0 & f - eb/a & g
 \end{gmatrix}.
 \]
 In this case, we count the following operations.
 First, there are $c-1$ divisions to compute the factors corresponding to $e/a$ in the above sketch, 
 where $c$ is the number of nonzero elements in the pivot column. 
 Secondly, there are $(r-1)(c-1)$ multiplications to compute all the numbers corresponding to $eb/a$, 
 where $r$ is the number of nonzero elements in the pivot row.
 Finally, for each clash of two nonzero elements in the submatrix, like $f$ and $eb/a$ in the sketch above,
 we count one addition.

 The second variant is inspired by fraction free elimination~\cite{geddes92}. Here we do not compute a multiplicative inverse
 of the pivot. Instead, the affected rows in the submatrix are multiplied by the pivot:
 \[
 \begin{gmatrix}[p]
   a & b & 0 \\
   0 & c & d \\
   e & f & g
   \rowops
   \add[-e]02
 \end{gmatrix}\leadsto
 \begin{gmatrix}[p]
   a & b & 0 \\
   0 & c & d \\
   0 & af - eb & ag
 \end{gmatrix}.
 \]
 In this case, we count the following operations.
 First, the number of multiplications by $a$ is given by $\sum_i (r_i-1)$ where $r_i$ is the number of nonzero entries in the $i$th
 row and the summation ranges over the rows which have a nonzero entry in the pivot column, excluding the pivot row.
 Secondly, there are again $(r-1)(c-1)$ multiplications compute all the numbers corresponding to $eb$ in the above sketch,
 where $r$ is is the number of nonzero elements in the pivot row and $c$ the number of nonzero elements
 in the pivot column.
 Finally, for each clash of two nonzero elements in the submatrix, like $af$ and $eb$ in the sketch above,
 we count one addition.
 
 In the following, we refer to the first variant as the ``field case'' (because it involves a division) and to the second variant
 as the ``ring case'' (because it is fraction free). 
 
 \section{How Many Matrices Are There?}

 It is clear that when we distinguish two kinds of entries, $0$~and~$\ast$, then
 there are $2^{n^2}$ different matrices of size $n\times n$.
 However, for the problem under consideration, we do not need to consider all of them.
 Pivot search will not be affected by permuting the rows of a matrix.
 For two $n\times n$ matrices $A,B$, write $A\approx B$ if a suitable permutation of
 rows turns $A$ into~$B$.
 Then $\approx$ is an equivalence relation, and we get a normal form with respect to $\approx$
 by simply sorting the rows. It suffices to consider these normal forms,
 which reduces the problem from $2^{n^2}$ matrices to $\binom{2^n+n-1}n$ equivalence classes.
 The number of equivalence classes is the number of sorted $n$-tuples of binary numbers less than~$2^n$.

 As we consider a pivot search that also allows column exchanges, we can go a step
 further. Write $A\sim B$ if applying a suitable permutation to the rows and a suitable
 permutation to the columns turns $A$ into~$B$.
 This is also an equivalence relation, but it is less obvious how to get a normal form.
 It was observed by Zivkovic~\cite{Zi:COSM} that deciding $\sim$ on
 binary matrices is equivalent to deciding the graph isomorphism problem for bipartite graphs.
 The idea is to interpret the matrix as adjacency matrix where row indices correspond to vertices
 of the first kind and column indices correspond to vertices of the second kind of a bipartite
 graph. Permuting rows or columns then amounts to applying permutations to each of the two
 kinds of vertices. Graph isomorphism is a difficult problem, but it is not a big deal for
 the matrix sizes we consider here. We used the nauty library~\cite{Mc:PGI} to compute normal
 forms with respect to~$\sim$, and only analyzed matrices in normal form.

 There does not appear to be a simple formula for the number of equivalence classes with
 respect to~$\sim$, but for small~$n$, the numbers can be determined by Polya
 enumeration theory, as explained for example in~\cite{HP:GE}, and they are available
 as A002724 in the OEIS~\cite{OEIS}.
 Here are the counts for $n=1,\dots,9$. 
 			
 \begin{center}
		\begin{tabular}{|r|r|r|r|}
			\hline
			\multicolumn{1}{|c|}{$n$} & \multicolumn{1}{c|}{$|S^{n\times n}|$} & \multicolumn{1}{c|}{$|S^{n\times n}/{\approx}|$} &  \multicolumn{1}{c|}{$|S^{n\times n}/{\sim}|$}\\
			\hline 
			1 & 2 & 2 & 2 \\
			\hline 
			2 & 16 & 10 & 7 \\
			\hline 
			3 & 512 & 120 & 36 \\
			\hline 
			4 & \numprint{65536} & \numprint{3876} & 317 \\
			\hline 
			5 & \numprint{33554432} & \numprint{376992} & \numprint{5624} \\
			\hline 
			6 & \numprint{68719476736} & \numprint{119877472} & \numprint{251610} \\
			\hline 
			7 & \numprint{562949953421312} & \numprint{131254487936} & \numprint{33642660} \\
			\hline 
			8 & \numprint{18446744073709551616} & \numprint{509850594887712} & \numprint{14685630688} \\
			\hline
			9 & \numprint{2417851639229258349412352} & \numprint{7145544812472168960} & \numprint{21467043671008} \\
			\hline
		\end{tabular}
             \end{center}

 An $8\times8$ binary matrix can be conveniently represented in a 64bit word,
 and for storing some information about the elimination cost for various pivot
 choices (best, worst, average, Markowitz), we spend altogether 56 bytes per
 matrix. With this encoding, a database for all $6\times 6$ matrices consumes
 about 3.8 terabyte of space, a database for all $7\times 7$ matrices up to
 row permutations consumes about 7.3 terabyte, and a database for all $8\times8$
 matrices up to row and column permutations consumes 784 gigabyte. The number of
 equivalence classes for $9\times9$ matrices is so much larger that we have not considered
 them. Not only would a database for $9\times9$ cost an absurd amount of disk space,
 it would also make the programming more cumbersome and the analysis less efficient
 because we need more than one word per matrix.

 \section{Exhaustive Search}
 
 We use exhaustive search to create databases with the following information:
 the matrix itself, the cost of elimination over a field and over a ring with
 the best pivot and what is the best pivot, the cost with the worst pivot and
 what is the worst pivot, the median of the elimination costs, and the same
 considering only the pivots which produce minimum fill-in. For matrices of size
 $n\times n$ we only do one elimination step and then use the data from the
 database of matrices of size $n-1$. We include a simplified pseudocode to show
 how the data is produced. By $M(n)$ we denote a set of representatives of
 $S^{n\times n}$ with respect to~$\sim$. The function \texttt{Eliminate} takes a
 matrix and a row and column index and returns the result of performing
 one step of Gaussian elimination. The function \texttt{Cost} takes the same
 input and returns the number of operations needed
 in the elimination step. The mappings $\mathit{costmin}$, $\mathit{costmax}$
 and $\mathit{costmed}$ are the database entries.
		
		\begin{algorithm}[H]
			\SetKwFunction{Eliminate}{Eliminate}
			\SetKwFunction{Cost}{Cost}
			\SetKwFunction{Median}{Median}
			\SetKwFunction{Min}{Min}
			\SetKwFunction{Max}{Max}
			\SetKwInOut{Input}{input}\SetKwInOut{Output}{output}
			\Input{A size $n$, a set of $n \times n$ matrices $M(n)$ and three mappings $costmin$, $costmax$ and $costmed$ from $M(n-1)$ to the reals}
			\Output{three mappings $costmin$, $costmax$ and $costmed$ from $M(n)$ to the reals}
			\For{$m \in M(n)$}{
				\eIf{there exist $i,j \in \{1,\ldots,n\}$ with $\Cost(m,i,j)=0$}{
					$\overline{m} \leftarrow \Eliminate(m, i, j)$\\
					$costmin(m) \leftarrow costmin(\overline{m})$\\
					$costmax(m) \leftarrow costmax(\overline{m})$\\
					$costmed(m) \leftarrow costmed(\overline{m})$
				}{
				$costmin(m) \leftarrow \infty;costmax(m) \leftarrow -\infty$\\
				\For{\normalfont \textbf{all} $(i,j)\in\{1,\dots,n\}^2$ \textbf{with} $m_{i j} = \ast$}{
							$\overline{m} \leftarrow \Eliminate(m, i, j)$\\
							$o \leftarrow \Cost(m, i, j)$\\
							$costmin(m) \leftarrow \Min(costmin(\overline{m})+o, costmin(m))$\\
							$costmax(m) \leftarrow \Max(costmax(\overline{m})+o, costmax(m))$\\
							$cost_{i j} \leftarrow costmed(\overline{m}) + o$
				}
				$costmed(m) \leftarrow \Median(cost)$
				}
			}
		\end{algorithm}
		
		The median in line~15 is taken only over those pivots which have been considered. For the minimum fill-in strategy we adjust the \textbf{if} statement in line~9 such that only pivots which produce minimum fill-in are considered. In this case also the database entries are taken from the minimum fill-in strategy. 
		
		Note that not only a pivot with no other elements in its column results in zero elimination cost (since there is nothing to eliminate), but also a pivot with no other elements in its row results in no elimination cost, since it does not change any matrix entries apart from those which become $0$. In line~2 we ensure that regardless of the strategy we are analyzing a free pivot is always chosen. Also note that if there are several such pivots, then the order in which we chose them does not affect the total cost. 
		
		We use the function \emph{genbg} from the nauty package to produce the list of $n \times n$ matrices and we use the nauty package main function to compute a canonical form after the elimination step. 
		
		The full source code and the databases up to size $6\times 6$ can be found at https://github.com/jakobmoosbauer/pivots. The larger databases can be provided by the authors upon request.
				
		\section{Machine Learning}
                
			There have been some recent advances in applying machine learning for improving selection heuristics in symbolic computation. As it is the case for Gaussian elimination, many algorithms allow different choices which do not affect correctness of the result, however may have a large impact on the performance. Machine learning models have for example been applied in cylindrical algebraic decomposition~\cite{HE:AMLT} and Buchberger's algorithm~\cite{PS:LSSI}. For more applications see~\cite{En:MLFM}. 

                        Without going into too much detail on the background of machine learning, we give here a summary of our approach,
                        mainly in order to document the computations we performed in order to facilitate a proper interpretation of the
                        experimental results reported later. For explanations of the technical terms used in this section, we refer to the
                        literature on machine learning. 
                        
			We train a reinforcement learning agent~\cite{SB:RLAI} to select a good pivot in Gaussian elimination. In reinforcement learning an agent interacts with an environment by choosing an action in a given state. In every time step the agent is presented with a state~$s_t$, chooses an action $a_t$ and the environment returns the new state $s_{t+1}$ and a reward~$r_{t+1}$. In our case the state is the matrix and the action is a row and a column index. The new state is the matrix we get after performing the elimination with the chosen pivot and the reward is minus the number of operations needed in the elimination step. An episode is a sequence of steps that transform a matrix to row echelon form. The agent tries to maximize the return $G_t = \sum_{k=0}^{T-t} \gamma^k r_{t+k+1}$ with the discount factor $0<\gamma \leq 1$. The return measures the expected future rewards and the discount factor makes the agent prefer earlier rewards over later rewards. Since we have a bound on the length of the episode we can choose $\gamma = 1$, in this case the return equals the total number of operations needed. 
			
			A difference to the more common concept of supervised learning is that we do not need a training set with training data. Instead we can sample from all possible inputs and the reward signal replaces the training data. This reduces the problem that an agent can produce bad answers outside of the training pool. We sample matrices equally distributed from all binary matrices. During the learning process actions are chosen using an $\epsilon$-greedy policy. This means that with a probability of $\epsilon$ we choose a random action, with probability $1-\epsilon$ we choose a greedy action, i.e., the action the agent would choose in the given situation. This policy ensures accurate predictions for the actions the agent eventually chooses and also that we do not miss good choices because we never try them. 
			
			The main component of the reinforcement learning agent is the deep Q-network which approximates the Q-value function~\cite{MK:HLCT,SB:RLAI}. The Q-value function maps a state and an action to the expected return value. We can use this information to pick a pivot that has the lowest expected cost. Since we need to fix the network size in advance we can only consider matrices of bounded size. For the present paper we stick to small matrix sizes since we can evaluate the overall performance using the database and the network stays of easy manageable size.

We encode the row and column of the pivot by a one-hot vector. This means we use $2n$ inputs nodes which each correspond to a row or a column and set those of the pivot to~$1$ and the others to~$0$. We also tried to use the fill-in as additional input feature to see whether it improves the performance. After experimenting with different network structures we settled with a fully connected network with $n(n+2)$, respectively $n(n+2)+1$ input nodes and two hidden layers with $n(n+2)$ nodes and a relu activation function. We chose this architecture to fit the task at hand, for larger matrices training and evaluating this neural network would get prohibitively expensive. Selecting appropriate features or decomposing the problem in smaller parts are possible ways around this.
				
			The network is trained using deep Q-learning with experience replay and a target network \cite{MK:HLCT}. Experience replay means that we keep a replay buffer with state-action-reward pairs we created and in each training step we sample a batch of training data for the network to learn from. This helps the network not to ``forget'' what it already learned. 
		
				For the learning process there are different hyperparameters which control the learning process. They can have a huge impact on the performance and the convergence of the learning process. We did not invest a lot of time in tuning network architecture and hyperparameters, since our goal was to evaluate the general applicability of machine learning to this problem rather than finding best-possible results. The $\epsilon$-greedy policy starts at $\epsilon =  0.5$ and slowly decays to $\epsilon = 0.1$. We use a discount factor of~$1$, learning rate~$0.001$, and a batch size of~50. The learning rate describes the step size of the parameter adjustment for the neural network. If the learning rate is too small, then the convergence is very slow, if the learning rate is too big the process does not converge at all.
							
			The goal of reinforcement learning to maximize a reward function directly applies to our problem, which is to minimize the number of operations needed. Another advantage is that we have a totally observable deterministic environment. A difficulty we are faced with is that we can choose among a large number of different actions, which depends on the current state. So for each possible action we need an extra call to the neural network. This could be avoided by only considering pivots with small fill-in instead of all possible pivots. Neural networks are quite good at pattern recognition, which seems to be useful in our context. However, since the computational cost is invariant under row and column permutation, there is no locality of features in the pivot selection problem. Therefore it is not reasonable to use convolutional layers, which proved very useful in other tasks involving pattern recognition. 

                        \vfill
         \section{Results}
                                
In this section we analyze the results of the exhaustive search and the
performance of machine learning. In the first graph we show the savings
achieved by using the minimum fill-in strategy compared to the median cost. We
observe that for matrices of size $8 \times 8$ the Markowitz criterion saves on average 42\% of the
operations needed in the field case and 37\% in the ring case (left graph).
Moreover, these numbers tend to grow as the matrices grow larger. So it is
reasonable to expect that even larger savings are achieved for very large matrices.

In the graph on the right we compare the median cost when choosing pivots with
minimal fill-in to the optimal pivot. For matrices of size $6 \times 6$ to $8
\times 8$ there are possible savings of about 5\% in the field case and about
7\% in the ring case. These numbers are increasing up to size $7\times 7$, but
there is a decrease from $7\times 7$ to $8 \times 8$. In view of this decrease,
it is hard to predict how the graph continues for matrix sizes that are
currently out of the reach of exhaustive search.

                \begin{center}\scriptsize
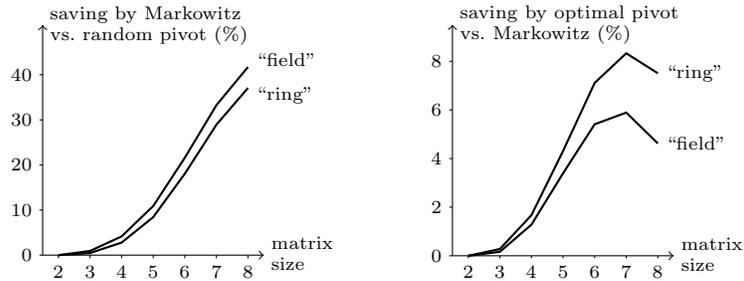
                \begin{tikzpicture}[scale=.6, xscale=.7, yscale=.1]
                  \draw[->] (1.5,0)--(1.5,51) node[right] {\vbox{\rlap{\strut saving by Markowitz}\rlap{\strut vs. random pivot (\%)}}};
                  \draw[->] (1.5,0)--(8.5,0) node[right] {\vbox{\rlap{\strut matrix}\rlap{\strut size}}};
                  \foreach \x in {2,...,8} \draw (\x,0)--(\x,-.7) node[below] {\x};
                  \foreach \y in {0,10,...,40} \draw (1.5,\y)--(1.4,\y) node[left] {\y};
                  \draw[thick] (2,0)--(3,0.925926)--(4,4.15459)--(5,10.894)--(6,21.5989)--(7,33.2554)--(8,41.7243) node[right,yshift=1mm] {``field''};
                  \draw[thick] (2,0)--(3,0.462963)--(4,2.78425)--(5,8.43237)--(6,18.0534)--(7,28.9232)--(8,37.0764) node[right,yshift=-1mm] {``ring''};
                \end{tikzpicture}
                \hfil
                \begin{tikzpicture}[scale=.6,scale=.7, yscale=.77]
                  \draw[->] (1.5,0)--(1.5,9.5) node[right] {\vbox{\rlap{\strut saving by optimal pivot}\rlap{\strut vs. Markowitz (\%)}}};
                  \draw[->] (1.5,0)--(8.5,0) node[right] {\vbox{\rlap{\strut matrix}\rlap{\strut size}}};
                  \foreach \x in {2,...,8} \draw (\x,0)--(\x,-.1) node[below] {\x};
                  \foreach \y in {0,2,...,8} \draw (1.5,\y)--(1.4,\y) node[left] {\y};
                  \draw[thick] (2,0)--(3,0.163399)--(4,1.27942)--(5,3.39172)--(6,5.40887)--(7,5.89114)--(8,4.62224) node[right] {``field''};
                  \draw[thick] (2,0)--(3,0.277778)--(4,1.67062)--(5,4.30875)--(6,7.1093)--(7,8.33055)--(8,7.5082) node[right] {``ring''};
                \end{tikzpicture}
                \end{center}
                
            For matrices of size $6 \times 6$ we also analyzed how the improvement potential depends on the sparsity of the matrix. For matrices with less than 30\% nonzero entries, almost all choices of pivots perform equally. We see that the highest savings compared to the minimum fill-in strategy can be made for matrices with 40 to 60\% nonzero entries, in the ring case up to 80\% nonzero entries. Although for rather dense matrices there is still room for sparsity improvements, for these matrices the Markowitz strategy is almost optimal. During the elimination matrices become denser every step, as we introduce new nonzero entries. By minimizing the fill-in we look ahead only one step in the elimination process. For matrices which are almost dense this seems to be sufficient, whereas for sparser matrices it might be helpful to look ahead further. 
                
                \begin{center}\scriptsize
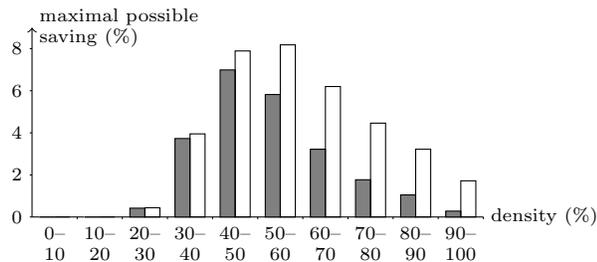
                  \begin{tikzpicture}[scale=.4,xscale=1.5,yscale=.7]
                  \draw[->] (0,0)--(0,9) node[right] {\vbox{\rlap{\strut maximal possible}\rlap{\strut saving (\%)}}};
                  \draw (0,0)--(10,0) node[right] {\rlap{density (\%)}};
                  \foreach \p/\x in {/1,1/2,2/3,3/4,4/5,5/6,6/7,7/8,8/9,9/10} \draw (\x,0)--(\x,-.1) ++(-.5,0) node[below] {\vbox{\clap{\p0--}\clap{\x0}}};
                  \foreach \y in {0,2,...,8} \draw (0,\y)--(-.05,\y) node[left] {\y};
                  \foreach \x/\p in {0/0,1/0,2/.42,3/3.73,4/6.99,5/5.82,6/3.22,7/1.77,8/1.05,9/.28}
                     \draw[fill=gray] (\x,0) ++ (.167,0) rectangle ++(.333,\p); 
                  \foreach \x/\p in {0/0,1/0,2/.44,3/3.95,4/7.89,5/8.18,6/6.2,7/4.46,8/3.22,9/1.72}
                     \draw (\x,0) ++ (.167,0) ++(.333,0) rectangle ++(.333,\p); 
                  \end{tikzpicture}
                \end{center}
                
            Since several different pivots can have minimal fill-in, the question arises whether we just need a refined criterion to pick an optimal pivot among those that have minimal fill-in, or if we need to do something completely different. In order to address this question, we test if the best pivot which produces minimal fill-in is already optimal. We observe that the percentage of matrices where Markowitz's strategy is optimal throughout the whole computation drops to 46\% in the field case and 30\% in the ring case for $8\times 8$ matrices. It seems reasonable to assume that for large matrices there is almost always a better strategy, even if you could choose the best pivot among those which have minimal fill-in.		
		Even though pivots that have minimal fill-in are not always optimal, it seems reasonable that optimal pivots still have rather small fill-in. Although we did not analyze this with our database, we did not observe any examples where the optimal pivot had very large fill-in compared to other choices. 

            	\begin{center}\scriptsize       
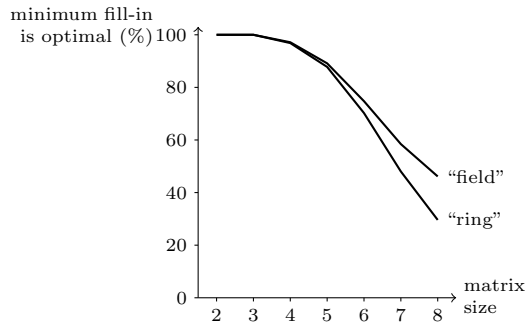
                \begin{tikzpicture}[scale=.7,xscale=.7,yscale=.05]
                  \draw[->] (1.5,0)--(1.5,103) node[left,xshift=-.5cm] {\vbox{\llap{\strut minimum fill-in}\llap{\strut is optimal (\%)}}};
                  \draw[->] (1.5,0)--(8.5,0) node[right] {\vbox{\rlap{\strut matrix}\rlap{\strut size}}};
                  \foreach \x in {2,...,8} \draw (\x,0)--(\x,-1.4) node[below] {\x};
                  \foreach \y in {0,20,...,100} \draw (1.5,\y)--(1.4,\y) node[left] {\y};
                  \draw[thick,yshift=30cm] (2,70)--(3,70)--(4,67.16)--(5,59.14)--(6,44.75)--(7,28.44)--(8,16.14) node[right] {``field''};
                  \draw[thick,yshift=30cm] (2,70)--(3,70)--(4,66.85)--(5,57.78)--(6,40.23)--(7,18.02)--(8,-.4) node[right] {``ring''};
                \end{tikzpicture}
                \end{center}

        The table below shows the improvement achieved by the machine learning model compared to the fill-in strategy. The machine learning model is able to surpass the Markowitz strategy by a very small amount. The network was trained on 40000 to 100000 matrices to a point where additional training did not result in further improvement. While for $4\times 4$ and $5 \times 5$ matrices this means that the model was presented every matrix multiple times, for the larger sizes the model was only trained on a small part of all matrices. This indicates that the model can generalize well from a small amount of samples. The results achieved using the fill-in as additional input to the network did not noticeably differ from those where we did not provide it. However, using the fill-in as feature we needed fewer training episodes to achieve similar results. So the machine learning model was able to find a better strategy knowing only the matrix entries, but providing additional features helps to speed up training. 
        Since neither further training nor using a deeper network did improve the results it is likely that the model gets stuck in a local maximum.
        
			 \begin{center}
			   \begin{tabular}{l|c|c|c|c}
                             $n$ & 4 & 5 & 6 & 7 \\\hline
                             field & 0.75\% & 1.31\% & 1.72\% & 1.92\% \\
                             ring & 1.19\% & 2.19\% & 3.27\% & 3.99\%
			   \end{tabular}
             \end{center}        
        
		Let us consider an example where we can see why the minimum fill-in strategy does sometimes not perform very well. This is the $6 \times 6$ matrix (actually $6\times 5$, since the last row consists of zeros) with the largest difference between the best pivot and the best pivot that produces minimum fill-in both over a ring and over a field:
		\begin{equation*}
			\begin{pmatrix}
			\ast & \ast & \ast & \ast & \ast & \ast \\
			\ast & \ast & \ast & \ast & \ast & \ast \\
			0 & 0 & \ast & \ast & \ast & \ast \\
			0 & 0 & 0 & \ast & \ast & \ast \\
			0 & 0 & 0 & 0 & \ast & \ast
			\end{pmatrix}.
		\end{equation*}
		When the rows and columns are sorted like this it is easy to spot that if we choose a pivot in the first column, we get a row echelon form after one elimination step with 11 operations. However, the pivots in the first column produce fill-in~5 and those in the last row produce only fill-in~4. Choosing the pivot with the minimal fill-in results in four elimination steps and in each step we have to eliminate every row, resulting in a total of 32 field operations. Over a ring the difference becomes even larger, 15~operations are needed with the best pivot and 55 if we always pick what produces minimal fill-in. There are two takeaways from this example. First we notice that the fill-in for the better pivot is produced by elements in its row and the fill-in for the other pivot is produced by the column. Especially in the ring case this leads to a large amount of extra computations since for every element we want to eliminate all the elements in the corresponding row have to be multiplied by the pivot. This suggests to use some kind of weighted fill-in where the number of elements in the column is weighted higher than the number of elements in the row. Another heuristic criterion motivated by this example is to do a two-step lookahead. In the first two columns there is a block of four nonzero elements and all other elements are~0. If we choose one of these four elements as pivot, then we only have to eliminate one row and the elimination is completed for both rows, since the second column contains no other elements. 
		In this particular example the neural network finds the optimal pivot with the higher fill-in.

		\bibliography{pivots}

\begin{thebibliography}{10}
\providecommand{\url}[1]{\texttt{#1}}
\providecommand{\urlprefix}{URL }
\providecommand{\doi}[1]{https://doi.org/#1}

\bibitem{BR:CMT}
Brualdi, R.A., Ryser, H.J., et~al.: Combinatorial matrix theory, vol.~39.
  Springer (1991)

\bibitem{CT:TBEP}
Corless, R.M., Thornton, S.E.: The bohemian eigenvalue project. ACM
  Communications in Computer Algebra  \textbf{50}(4),  158–160 (February
  2017). \doi{10.1145/3055282.3055289}

\bibitem{DE:DMFS}
Duff, I.S., Erisman, A.M., Reid, J.K.: Direct methods for sparse matrices.
  Clarendon Pr (1986)

\bibitem{En:MLFM}
England, M.: Machine learning for mathematical software. In: Mathematical
  Software -- ICMS 2018. pp. 165--174. Springer International Publishing, Cham
  (2018)

\bibitem{faugere99}
Faug{\`e}re, J.C.: A new efficient algorithm for computing {G}r{\"o}bner bases.
  Journal of Pure and Applied Algebra  \textbf{139}(1--3),  61--88 (1999)

\bibitem{geddes92}
Geddes, K.O., Czapor, S.R., Labahn, G.: Algorithms for Computer Algebra. Kluwer
  (1992)

\bibitem{HP:GE}
Harary, F., Palmer, E.M.: Graphical enumeration. Acad. Press (1973)

\bibitem{HE:AMLT}
Huang, Z., England, M., Wilson, D., Davenport, J.H., Paulson, L.C., Bridge, J.:
  Applying machine learning to the problem of choosing a heuristic to select
  the variable ordering for cylindrical algebraic decomposition. In:
  Intelligent Computer Mathematics. pp. 92--107. Springer International
  Publishing, Cham (2014)

\bibitem{koutschan13}
Koutschan, C.: Creative telescoping for holonomic functions. In: Computer
  Algebra in Quantum Field Theory: Integration, Summation and Special
  Functions. pp. 171--194. Texts and Monographs in Symbolic Computation,
  Springer (2013)

\bibitem{Ma:TEFO}
Markowitz, H.M.: The elimination form of the inverse and its application to
  linear programming. Management Science  \textbf{3}(3),  255--269 (1957),
  \url{http://www.jstor.org/stable/2627454}

\bibitem{Mc:PGI}
McKay, B.D., Piperno, A.: Practical graph isomorphism, ii. Journal of Symbolic
  Computation  \textbf{60}(0),  94--112 (2014). \doi{10.1016/j.jsc.2013.09.003}

\bibitem{MK:HLCT}
Mnih, V., Kavukcuoglu, K., Silver, D., Rusu, A.A., Veness, J., Bellemare, M.G.,
  Graves, A., Riedmiller, M., Fidjeland, A.K., Ostrovski, G., et~al.:
  Human-level control through deep reinforcement learning. Nature
  \textbf{518}(7540),  529--533 (2015)

\bibitem{PS:LSSI}
Peifer, D., Stillman, M., Halpern-Leistner, D.: Learning selection strategies
  in buchberger's algorithm. arXiv preprint arXiv:2005.01917  (2020)

\bibitem{OEIS}
Sloane, N.J.A.: The on-line encyclopedia of integer sequences (2020),
  \url{https://oeis.org/}

\bibitem{SB:RLAI}
Sutton, R.S., Barto, A.G.: Reinforcement learning: An introduction. MIT press
  (1998)

\bibitem{Ya:CTMF}
Yannakakis, M.: Computing the minimum fill-in is {NP}-complete. SIAM J.
  Algebraic Discrete Methods  \textbf{2}(1),  77--79 (1981).
  \doi{10.1137/0602010}

\bibitem{Zi:COSM}
Živković, M.: Classification of small (0,1) matrices. Linear Algebra and its
  Applications  \textbf{414}(1),  310--346 (2006).
  \doi{10.1016/j.laa.2005.10.010}

\end{thebibliography}
\end{document}